\documentclass[aps,prl,twocolumn, superscriptaddress,showpacs,preprintnumbers,amsmath,amssymb]{revtex4}
\usepackage{color}
\usepackage{longtable}
\usepackage{graphicx} 
\usepackage{dcolumn}  

\graphicspath{{ps}}

\begin{document}


 
\title{ \quad\\[0.5cm] Observation of $X(3872)\to J/\psi \gamma$  and 
search for $X(3872)\to\psi'\gamma$ in $B$ decays}

\affiliation{Budker Institute of Nuclear Physics, Novosibirsk}
\affiliation{Faculty of Mathematics and Physics, Charles University, Prague}
\affiliation{University of Cincinnati, Cincinnati, Ohio 45221}
\affiliation{Justus-Liebig-Universit\"at Gie\ss{}en, Gie\ss{}en}
\affiliation{Gyeongsang National University, Chinju}
\affiliation{Hanyang University, Seoul}
\affiliation{University of Hawaii, Honolulu, Hawaii 96822}
\affiliation{High Energy Accelerator Research Organization (KEK), Tsukuba}
\affiliation{Indian Institute of Technology Guwahati, Guwahati}
\affiliation{Indian Institute of Technology Madras, Madras}
\affiliation{Indiana University, Bloomington, Indiana 47408}
\affiliation{Institute of High Energy Physics, Chinese Academy of Sciences, Beijing}
\affiliation{Institute of High Energy Physics, Vienna}
\affiliation{Institute of High Energy Physics, Protvino}
\affiliation{Institute for Theoretical and Experimental Physics, Moscow}
\affiliation{J. Stefan Institute, Ljubljana}
\affiliation{Kanagawa University, Yokohama}
\affiliation{Institut f\"ur Experimentelle Kernphysik, Karlsruher Institut f\"ur Technologie, Karlsruhe}
\affiliation{Korea Institute of Science and Technology Information, Daejeon}
\affiliation{Korea University, Seoul}
\affiliation{Kyungpook National University, Taegu}
\affiliation{\'Ecole Polytechnique F\'ed\'erale de Lausanne (EPFL), Lausanne}
\affiliation{Faculty of Mathematics and Physics, University of Ljubljana, Ljubljana}
\affiliation{University of Maribor, Maribor}
\affiliation{Max-Planck-Institut f\"ur Physik, M\"unchen}
\affiliation{University of Melbourne, School of Physics, Victoria 3010}
\affiliation{Nagoya University, Nagoya}
\affiliation{Nara Women's University, Nara}
\affiliation{National Central University, Chung-li}
\affiliation{Department of Physics, National Taiwan University, Taipei}
\affiliation{H. Niewodniczanski Institute of Nuclear Physics, Krakow}
\affiliation{Nippon Dental University, Niigata}
\affiliation{Niigata University, Niigata}
\affiliation{University of Nova Gorica, Nova Gorica}
\affiliation{Novosibirsk State University, Novosibirsk}
\affiliation{Osaka City University, Osaka}
\affiliation{Pacific Northwest National Laboratory, Richland, Washington 99352}
\affiliation{Panjab University, Chandigarh}
\affiliation{Research Center for Nuclear Physics, Osaka}
\affiliation{University of Science and Technology of China, Hefei}
\affiliation{Seoul National University, Seoul}
\affiliation{Sungkyunkwan University, Suwon}
\affiliation{School of Physics, University of Sydney, NSW 2006}
\affiliation{Tata Institute of Fundamental Research, Mumbai}
\affiliation{Excellence Cluster Universe, Technische Universit\"at M\"unchen, Garching}
\affiliation{Tohoku Gakuin University, Tagajo}
\affiliation{Tohoku University, Sendai}
\affiliation{Department of Physics, University of Tokyo, Tokyo}
\affiliation{Tokyo Institute of Technology, Tokyo}
\affiliation{Tokyo Metropolitan University, Tokyo}
\affiliation{Tokyo University of Agriculture and Technology, Tokyo}
\affiliation{CNP, Virginia Polytechnic Institute and State University, Blacksburg, Virginia 24061}
\affiliation{Wayne State University, Detroit, Michigan 48202}
\affiliation{Yonsei University, Seoul}

  \author{V.~Bhardwaj}\affiliation{Panjab University, Chandigarh} 
  \author{K.~Trabelsi}\affiliation{High Energy Accelerator Research Organization (KEK), Tsukuba} 
  \author{J.~B.~Singh}\affiliation{Panjab University, Chandigarh} 
 \author{S.-K.~Choi}\affiliation{Gyeongsang National University, Chinju} 
 \author{S.~L.~Olsen}\affiliation{Seoul National University, Seoul}\affiliation{University of Hawaii, Honolulu, Hawaii 96822} 

  \author{I.~Adachi}\affiliation{High Energy Accelerator Research Organization (KEK), Tsukuba} 
 \author{K.~Adamczyk}\affiliation{H. Niewodniczanski Institute of Nuclear Physics, Krakow} 
  \author{D.~M.~Asner}\affiliation{Pacific Northwest National Laboratory, Richland, Washington 99352} 
  \author{V.~Aulchenko}\affiliation{Budker Institute of Nuclear Physics, Novosibirsk}\affiliation{Novosibirsk State University, Novosibirsk} 
  \author{T.~Aushev}\affiliation{Institute for Theoretical and Experimental Physics, Moscow} 
% ...
\collaboration{The Belle Collaboration}


\begin{abstract}
We report  a study of  $B\to  (J/\psi \gamma) K$ and  $B\to (\psi' \gamma)K$ decay modes using 
$772\times 10^{6}$ $B\overline{B}$ events collected at the $\Upsilon(4S)$ 
resonance with the Belle detector at the KEKB energy-asymmetric $e^+ e^-$ 
collider. We observe $X(3872) \to J/\psi \gamma$  and report the
 first evidence for
$\chi_{c2} \to J/\psi \gamma$  in $B\to (X_{c\overline{c}}\gamma)
 K$  decays, while in  a search for $X(3872) \to \psi' \gamma$  no 
significant signal is found. We measure the  branching fractions, 
$\mathcal{B}(B^{\pm} \to X(3872) K^{\pm})  \mathcal{B}(X(3872) 
\to J/\psi\gamma)$ $=$ $(1.78^{+0.48}_{-0.44}\pm 0.12)\times 10^{-6}$, $\mathcal{B} (B^{\pm} \to\chi_{c2} K^{\pm})$$=$ $(1.11^{+0.36}_{-0.34} \pm 0.09) 
\times 10^{-5}$,  $\mathcal{B}(B^{\pm} \to X(3872) K^{\pm})  \mathcal{B}(X(3872) 
\to \psi'\gamma)$ $<$ $3.45\times 10^{-6}$  (upper limit at 90\% C.L.) and also  provide upper limits  for other searches.

\end{abstract}

\maketitle

The $X(3872)$ state was observed by the Belle Collaboration~\cite{belle1} in 
2003, and later confirmed by CDF~\cite{cdf1}, D$\O$~\cite{do1} and 
BaBar~\cite{babar1}. The fact that it was not seen  in decays to 
$\chi_{c1}\gamma$, $\chi_{c2}\gamma$, and $J/\psi \eta$ final states 
 suggests that the $X(3872)$  is not a 
 conventional $q\bar{q}$ meson state 
 that  can be explained by  a simple quark 
model~\cite{belle1,bellep,babarprl93}. Because of its narrow width and 
 the proximity of its mass, $3871.5 \pm 0.2$ MeV$/c^2$~\cite{wavg} 
to the $D^{*0}\overline{D}{}^0$  
threshold,  the $X(3872)$ is a good candidate for a  $D\overline{D}{}^*$ 
 molecule~\cite{ericplb598}. Other possibilities have also been proposed for 
 the $X(3872)$ state, such as tetraquark \cite{Lmaiani}, $c\overline{c}g$ hybrid 
meson \cite{Lihybrid} and vector glueball  models \cite{kkseth}. 

Radiative decays of  the $X(3872)$ are important in understanding its nature. 
One such decay, $X(3872)\to J/\psi \gamma$, \cite{bellep,babarjg} established 
its charge parity to be +1.  In the molecular model, the radiative decays 
of  the $X(3872)$ occur through vector meson dominance (VMD) and light quark 
annihilation (LQA) \cite{ericplb598}. The  decay rate of 
$X(3872)\to J/\psi \gamma$ is dominated 
by VMD while for $X(3872)\to \psi'\gamma$
\cite{psi(2S)}, it is mostly driven by LQA, 
implying  that $X(3872)$ decay to $\psi'\gamma$  is highly suppressed compared 
to $J/\psi \gamma$ \cite{ericplb598}. Recent results from the BaBar 
 Collaboration~\cite{babarprl102} 
show that $\mathcal{B}(X(3872) \to \psi' \gamma)$ is almost three times  
that of $\mathcal{B}(X(3872) \to J/\psi \gamma)$,  which is inconsistent  
with  a pure $D^{*0}\overline{D}{}^0$ molecular model, and can be interpreted 
as   indicating a $c\overline{c}$-$D^{*0}\overline{D}{}^0$ admixture \cite{ericplb598, msuzuki}. 
 If  the $X(3872)$ is an admixture of  $\chi_{c1}'$ and 
a molecular state, and its production and radiative decays are 
mainly due to its $\chi_{c1}'$ component, then the 
$\psi' \gamma$ decay, a favored E1 transition of 
$\chi_{c1}'$, should  be significantly enhanced compared to the $J/\psi \gamma$ 
decay,  which is ``hindered" by poor wave function overlap \cite{barnesgodfray}.

In this Letter, we present  new results on $B \to (\chi_{c1},\chi_{c2}, 
X(3872)) K$, where  the $\chi_{c1},\chi_{c2},X(3872)$ decays to $J/\psi \gamma$, 
and  the $X(3872)$  decays to $\psi' \gamma$~\cite{mixchg}.   These results 
are obtained from the final data sample of $772\times 10^{6}$ $B\overline{B}$  
events collected  with the Belle detector~\cite{abashian} at the 
KEKB~\cite{kurokawa} energy-asymmetric $e^+e^-$ collider operating at 
the $\Upsilon(4S)$ resonance. The Belle detector is a large-solid-angle 
spectrometer which includes a silicon vertex detector, a 50-layer 
central drift chamber (CDC), an array of aerogel threshold Cherenkov 
counters (ACC), time-of-flight scintillation counters (TOF), and an 
electromagnetic calorimeter (ECL)  comprising CsI(Tl) crystals 
located inside a superconducting solenoid coil that provides a 1.5~T 
magnetic field.

The $J/\psi$ meson is reconstructed in its decays to $\ell^+\ell^-$ 
($\ell =$ $e$ or $\mu$), and the $\psi'$ meson in its decays to 
$\ell^+\ell^-$ and $J/\psi\pi^+ \pi^-$. 
In the $\psi'\to e^+e^-$ and $J/\psi \to e^+ e^-$ decays, 
the four-momenta of all photons within 50 mrad of each 
of the original $e^+$ or $e^-$ tracks are included in the invariant mass 
calculation  $[$hereafter denoted as  $M_{e^+e^- (\gamma)}]$,  in order to reduce
the radiative tail. The reconstructed invariant mass of the $J/\psi$ 
candidates is required to satisfy 2.95 GeV$/c^2 < M_{e^+ e^-(\gamma)} 
< 3.13$ GeV$/c^2$ or $3.03$ GeV$/c^2  < M_{\mu^+ \mu^-} < 3.13$ GeV$/c^2$.  
In  the $\psi' \to \ell^+\ell^-$ reconstruction, the invariant mass is  restricted to the range 
$3.63$ GeV$/c^2$ $<$ $M_{e^+  e^- (\gamma )} < 3.72$ GeV$/c^2$ or 
$3.65$ GeV$/c^2 < M_{\mu^+ \mu-} < 3.72$ GeV$/c^2$. 
To reconstruct $\psi'\to J/\psi \pi^+ \pi^-$ decays, 
$\Delta M = M_{\ell^+ \ell^- \pi^+ \pi^-} - M_{\ell^+\ell^-}$ should 
satisfy the condition $0.58$ GeV$/c^2 < \Delta M < 0.60$ GeV$/c^2$. 
In order to reduce the combinatorial background due to low-momentum pions,  
the invariant mass of the two pions  from 
the $\psi'$ decay, $M_{\pi^+\pi^-}$, is  required to be greater than 
$0.40$ GeV$/c^2$.  A mass- and vertex-constrained fit is performed to 
all the selected $J/\psi$ and $\psi'$ candidates  to 
improve their momentum resolution. 

The $\chi_{c1,c2}$  and the $X(3872)$ candidates are  formed by combining  the 
$J/\psi$ candidates with a photon. The photons are reconstructed from energy 
depositions in  the ECL and  are required to have  energies  (in the lab frame) greater 
than 270 (470) MeV for $\chi_{c1,c2}$ ($X(3872)$) 
reconstruction. In  a similar fashion, $X(3872)$ candidates decaying to 
$\psi' \gamma$ are reconstructed by combining $\psi'$ candidates with 
$\gamma$ candidates  with  energies greater than 100 MeV. 

Charged tracks are identified as  pion or  kaon candidates using information from 
the CDC ($dE/dx$), TOF, and ACC systems. 
The kaon identification efficiency is $88\%$ while the probability of 
 a pion  misidentified as a kaon  is $10\%$.  The pions  used in 
the reconstruction of  the $\psi'$  in the $J/\psi \pi^+\pi^-$  channel  have an
identification efficiency of 99\%  with a kaon to pion misidentification probability of 2\%. 
 Candidate $K_S^0$ mesons are reconstructed by combining two oppositely charged  tracks (with a pion mass assumed) with 
invariant mass lying between $[ 0.482, 0.514 ]$ GeV$/c^2$; the selected 
candidates are required to  satisfy the criteria given in detail in 
Ref.~\cite{goodks}.

To reconstruct the $B$ candidates, each  $J/\psi \gamma$ or $\psi' \gamma$ system 
is combined with a kaon candidate. Two kinematic variables are formed: the 
beam-constrained mass ($M_{\rm bc} \equiv \sqrt{{{E}^{*2}_{\rm beam}} - 
{p^{*2}_{B}}}$) and the energy difference ($\Delta E \equiv E_{B}^*- 
E^*_{\rm beam}$). Here  ${E^*_{ \rm beam}}$ is the run-dependent beam energy, 
and ${E^*_{B}}$ and ${p^*_{B}}$  are the reconstructed energy and momentum, 
respectively, of the $B$ meson candidates in  the  $\Upsilon(4S)$
 center-of-mass (CM) frame. Candidates having $M_{\rm bc} > 5.27$ 
GeV$/c^2$ and within a $\Delta E$ window 
of $[-25, 30]$ MeV for $\chi_{c1,c2}$ and $[-30, 35]$ MeV ($[-20, 20]$ MeV) 
for $X(3872)\to J/\psi \gamma$ ($X(3872) \to \psi' \gamma$) are  retained for further analysis. 
 We extract the signal yield by performing an unbinned extended 
maximum likelihood fit to the variable $M_{\psi \gamma}$ defined as 
$M_{\ell\ell \gamma} - M_{\ell\ell} + m_{\psi}$~\cite{psigam},
 where  $m_{\psi}$  is the world average mass~\cite{pdg10}. 
In order to improve the resolution of $M_{\psi \gamma}$, we scale the energy of
 the $\gamma$  so that $\Delta E$ is equal to zero. 

To suppress continuum background, events having a ratio of the second to zeroth 
Fox-Wolfram moments~\cite{foxwolfram} $R_2 > 0.5$ are rejected. 
Large $B\to \psi X$ MC samples (corresponding to $50$ times the data sample size 
used in this analysis) are used to study the background. 
 To study the non-$J/\psi$ (non-$\psi'$) background  $M_{\ell\ell}$ sidebands in data,
within [2.5-2.6] GeV/$c^2$ ([3.35-3.45] GeV$/c^2$) and [3.2-3.5] GeV/$c^2$ 
([3.8-4.0] GeV/$c^2$),  are used.

 For the $(J/\psi \gamma) K$  channels, the background   is primarily from 
$B \to J/\psi K^*$  decays   that do not peak in $M_{J/\psi \gamma}$. 
To reduce this background, we veto candidate photons from 
 $\pi^0 \to \gamma \gamma$  by combining them with 
 any other photon and  then by rejecting both $\gamma$'s in the pair if the 
 $\pi^0$ likelihood is greater than 0.52. 
This likelihood is a function of the laboratory energy 
of the other photon, its polar angle and   the invariant mass of the two-photon system, and 
 is determined using  MC study~\cite{kopenberg}.
We also reject  photon candidates  with $\cos \theta_{\rm hel} >0.76 
~(>0.85)$ in  the $\chi_{c1,c2}$ ($X(3872)$)  selection, where  the helicity angle $\theta_{\rm hel}$ 
is defined as the angle between the direction of the 
photon and the direction opposite to the $B$ momentum in the $\chi_{c1,c2}$ ($X(3872)$) rest frame. 
  Applying these  criteria,  the background is reduced by 86\% (79\%)  with a signal loss 
 of 35\% (30\%) for  the $B \to \chi_{c2} K$ ($B\to X(3872) K$) decay mode. 
 For  1.3\%  of  events with multiple candidates 
in $B \to (J/\psi \gamma ) K$ decay modes, 
we  select the $B$ candidate having $M_{\rm bc}$ closest to the nominal 
$B$ mass~\cite{pdg10}.

 A sum of two Gaussians is used to model  the signal shapes
of $B\to\chi_{c1}K$ and $B\to \chi_{c2}K$. The fraction of 
each Gaussian is fixed  to the value obtained  from MC simulated events. 
 For $B^+ \to \chi_{c1} K^+$  the other shape parameters are  floated
in the fit  whereas for $B^+ \to \chi_{c2} K^+$ they are fixed using 
the mass difference (from Ref.~\cite{pdg10}) and the width difference 
(from MC simulations) between  the $\chi_{c1}$ 
and $\chi_{c2}$.   The non-peaking combinatorial background component is modeled  with
a second-order polynomial. 
For  the $B^{0} \to \chi_{c1} K^0_S$ and $B^{0} \to \chi_{c2} K^0_S$ decay modes, the 
signal shape is fixed using the  results from the charged $B$ mode.

\begin{figure}[h!]
\begin{center}
\includegraphics[scale=0.45]{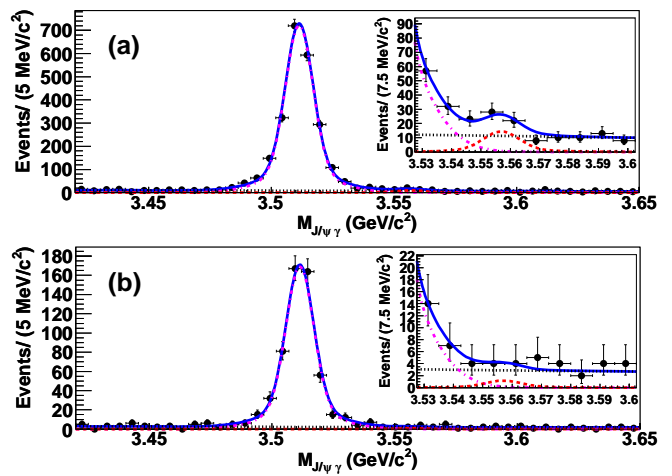}

\caption{\label{fig:chic1k} $M_{J/\psi\gamma}$ distributions for (a) 
 $B^{+} \to \chi_{c1,c2}(\to J/\psi\gamma)K^{+}$ and (b) 
 $B^{0}\to\chi_{c1,c2}(\to J/\psi\gamma)K^0_S$ 
decays. The curves show the signal (pink dot-dashed for  
$\chi_{c1}$ and red dashed 
for $\chi_{c2}$), and the background component (black dotted) as well as  the overall  fit (blue solid).  The  insets show a reduced range of $M_{J/\psi \gamma}$ and the contribution of the $B\to \chi_{c2}K$ peak.}
\end{center}
\end{figure}

Figure 1 shows the fit to  the $M_{J/\psi \gamma}$ distribution for 
$B \to \chi_{c1} K$  and $B \to \chi_{c2} K$ decays in the range of [3.38, 3.70] GeV$/c^2$.
We observe  the $\chi_{c1}$ in both $B$ decay modes, and obtain 3.6 
standard deviation ($\sigma$) evidence for  the $\chi_{c2}$ in the charged $B$ 
decay  mode. The statistical significance is defined as 
$\sqrt{-2 \ln (\mathcal{L}_0/ \mathcal{L}_{\rm max})}$ where 
$\mathcal{L}_{\rm max}$ ($\mathcal{L}_0$) denotes the likelihood value 
when the yield is allowed to vary (is  set to zero).   The systematic
uncertainty,  which is described below, is included in the significance \cite{rdcousins}. 
As no significant signal is found for $B^0 \to \chi_{c2} K^0$, 
we  determine a 90\% confidence level (C.L.) upper limit (U.L.) on its branching fraction  with
 a frequentist method  that uses ensembles of pseudo-experiments. 
For a given signal yield, 10000 sets of signal and background events are 
generated according to their PDFs, and fits are performed. The U.L. is  determined from 
 the fraction of samples that give a yield larger than that of data.

For the $B \to X(3872)(\to J/\psi \gamma) K$ decay mode, a sum of two Gaussians 
is  also used to model the signal PDF and the combinatorial background 
component is modeled by a first-order  polynomial. 
To take into account small differences between the MC simulation and data,
the  signal PDF shapes are corrected  for calibration factors 
determined  from  the $B^+ \to \chi_{c1} K^+$  fit. 
Figure 2 shows the fit to  the $M_{J/\psi \gamma}$ distributions for 
$B \to X(3872) K$  performed in the range [3.7, 4.1] GeV$/c^2$.
We  find a clear signal for $X(3872) \to J/\psi \gamma$ 
in the charged decay $B^+ \to X(3872) K^+$ with a
significance  of $4.9\sigma$  and measure the  product branching fraction 
$\mathcal{B} (B^+\to X(3872)K^+)  \mathcal{B} (X(3872) \to 
J/\psi \gamma)$ 
 $=$ $(1.78^{+0.48}_{-0.44} (\rm stat.) \pm 0.12 (\rm syst.)) \times 10^{-6}$. 
We also give  an U.L. on the branching fraction for the neutral $B$ mode whose significance is 2.4$\sigma$
(Table~\ref{tab_results}). We  estimate the significance of 
the $X(3872)\to J/\psi\gamma$ signal by simultaneously fitting the charged and 
the neutral $B$ decay modes; we obtain a significance of 5.5$\sigma$ including systematics  uncertainties. 

\begin{figure}[h!]
\begin{center}
\includegraphics[scale=0.45]{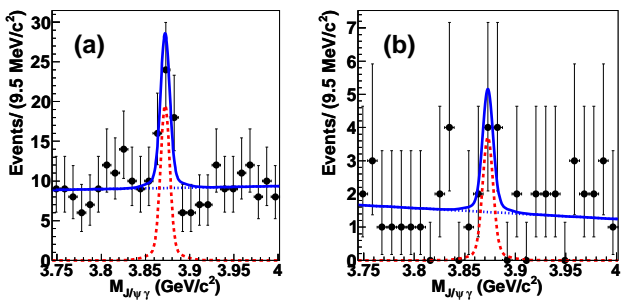}
\caption{\label{fig:x3_jk} $M_{J/\psi\gamma}$ distributions 
for (a)  $B^+\to X(3872)(\to J/\psi\gamma)K^+$ and (b) 
 $B^0\to X(3872)(\to J/\psi\gamma)K^0_S$ decays. The curves show the signal 
(red dashed) and the background component (blue dotted) as well as  the 
overall fit (blue solid).}
\end{center}
\end{figure}

For the $B\to(\psi' \gamma) K$ decay mode, the background has a broad peaking 
structure, most of which is from $B \to \psi' K^*$  decay mode. 
 Here, since the $\gamma$'s from $X(3872)\to\psi' \gamma$ have low
energy (less than one third of the energy
of the $\gamma$'s coming from $X(3872)\to J/\psi \gamma$),
 the $\pi^0$-veto and $\cos \theta_{\rm hel}$  selection result in more signal loss 
than background reduction. 
Instead, we combine the $\psi' K$ of the $\psi' \gamma K$ candidates 
with  any $\pi^{\pm}$ or  $\pi^0$ candidate in the event. 
Three variables, namely 
$\Delta E'$ ($\equiv E^*_{ \psi' } + E^*_{K^*} - E_{ \rm beam}^*$), 
$M_{\rm bc}'$ ($\equiv \sqrt{E^{*2}_{\rm beam} - 
(p^*_{\psi'} + p^*_{K^*})^2}$) and the invariant mass of $K\pi$ 
($M_{K\pi}$), are used for this purpose. 
Events satisfying the criteria of 817 MeV$/c^2 < M_{K\pi} < 967$ MeV$/c^2$, 
$\Delta E'$ within $[-20,20]$ MeV and $M_{\rm bc}' >$ 5.27 GeV$/c^2$, 
are identified as $B \to \psi' K^*$  candidates and discarded.
 This results in the reduction of the background by 59\% 
with a 22\% loss { of signal.
 For 15.4\%  of events with multiple candidates 
in $B \to (\psi' \gamma) K$ decay modes, 
we  select the $B$ candidate having $M_{\rm bc}$ closest to the nominal 
$B$ mass~\cite{pdg10}.

The branching fraction for the $B \to (\psi' \gamma) K$ mode is determined 
from a simultaneous fit performed to the two decay modes of the $\psi'$.   
The background shape for  $B \to (\psi' \gamma) K$  has both a peaking 
and a non-peaking component. For the peaking component, the shape is 
estimated from a large sample of MC simulated events of  
 $\psi' K$ and $\psi'K^*$, and their fractions are fixed 
using the branching  fractions from Ref.~\cite{pdg10}. 
The non-peaking background (combinatorial background) is parameterized by 
a threshold function $(M_{\psi'\gamma})^2 \times 
\exp(a ~(M_{\psi'\gamma} - M_{Th}) ~+~b~(M_{\psi' \gamma} - M_{Th})^2)$, 
where $M_{Th} = 3.725$ GeV$/c^2$.
 The $\psi$ mass data sidebands and  large $B \to \psi X$ MC 
sample (after removing $B\to \psi' K$ and $B \to \psi' K^*$ decays)
are used to estimate the parameters of the threshold function.
 The shapes for both background components are fixed whereas their  yields are 
 allowed to float in the fit. 
The signal is described as a sum of two Gaussians and is fixed from MC study 
after applying calibration corrections (from $B^+ \to \chi_{c1} K^+$ 
study) while its yield is  allowed to vary in the fit. 
 No significant bias is found in fitting  ensembles of
the simulated experiments containing  the signal and  background components.

Figure 3 shows the  results of the fit to  the $M_{\psi' \gamma}$ distribution for $B\to X(3872) K$.  
 The fitted yields are 
$5.0^{+11.9}_{-11.0}$  events ($1.5^{+4.8}_{-3.9} $  events) for  $B^+\to X(3872) K^+$ ($B^0 \to X(3872) K_S^0$). 
 Since there is no significant signal in either channel, we determine U.L.s  of $\mathcal{B}(B^+ \to X(3872) K^+)   \mathcal{B}(X(3872) \to \psi' \gamma)$ 
($\mathcal{B}(B^{0} \to X(3872) K^{0}) \mathcal{B}(X(3872) \to \psi' \gamma)$)  as  $ 3.45
 \times 10^{-6}$ ($ 6.62 \times 10^{-6}$) using the method described above. 
 A  completely independent analysis, with different selection  criteria and  a different fitting technique
was performed on the same data sample; the  results were found to be consistent with the results  reported  in this Letter.

\begin{figure}[h!]
\begin{center}
\includegraphics[scale=0.45]{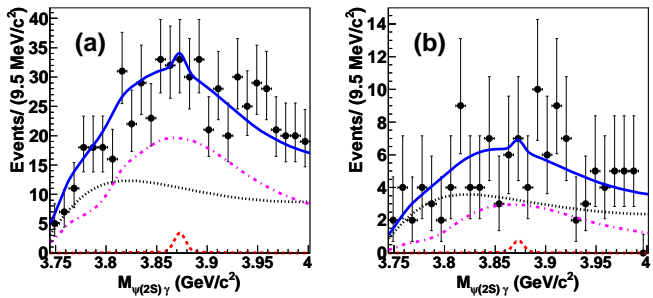}
\caption{\label{fig:xpsi2s} $M_{\psi' \gamma}$ distributions 
  for
 (a) $B^+\to$ $X(3872)$ $(\to \psi' \gamma)K^+$ and (b) $B^{0}\to$ $X(3872)$ $(\to \psi' \gamma)K^{0}$. 
The curves show the signal (red dashed for $X(3872)$) and  the background component (pink dot-dashed 
for background from $B\to \psi' K^*$ and $B\to \psi' K$ component, and black dotted for combinatorial background modeled by  the threshold function) as well as  the overall fit (blue solid). }
\end{center}
\end{figure}

 The branching fractions and the fit results are summarized in Table~\ref{tab_results}. 
Equal production of neutral and charged $B$ meson pairs in the $\Upsilon (4S)$ 
decay is assumed. Secondary branching fractions used to calculate 
$\mathcal{B}$ are taken from Ref.~\cite{pdg10}.

\begin{table}[h]
\caption{Corrected efficiency ($\epsilon$), signal 
yield ($Y$) from the fit,  measured $\mathcal{B}$  or 
90$\%$ C.L.  upper limit (U.L.) for  $B\to \chi_{c1,c2} K$, 
$B \to X(3872) (\to J/\psi\gamma) K$ and $B \to X(3872) (\to 
\psi'\gamma) K$ decay modes   and  significance ($\mathcal{S}$) with 
systematics included. $\mathcal{B}$ for $B\to X(3872) K$ is  the product
 $\mathcal{B} (B\to X(3872) K)  \mathcal{B} (X(3872)\to \psi \gamma)$. 
For $\mathcal{B}$, the first (second) error is statistical (systematic).   }
\begin{center}

\begin{tabular}{lccccc}
\hline \hline
Decay & $\epsilon$(\%) &  Yield ($Y$) &   Branching fraction & $\mathcal{S}~(\sigma)$ \\ \hline

\multicolumn{3}{c}{$B \to \chi_{c1}(\to J/\psi\gamma) K$~~~~~~~~~~~}  &  $\mathcal{B}$ ($\times 10^{-4}$) & \\ \hline
$K^{+}$ & 14.8  &  $2308_{-52}^{+53}$ & $4.94 \pm 0.11 \pm 0.33$&  79 \\
$K^0$ & 13.2 &  $542\pm24$ & $3.78_{-0.16}^{+0.17}\pm{0.33}$&  37 \\ \hline

\multicolumn{3}{c}{$B \to \chi_{c2}(\to J/\psi\gamma) K$~~~~~~~~~~~}  &  $\mathcal{B}$ ($\times 10^{-5}$) & \\ \hline
$K^{+}$ & 16.6 & $32.8_{-10.2}^{+10.9}$& $1.11_{-0.34}^{+0.36}\pm0.09$& $3.6$ \\
$K^0$ & 14.4 & $2.8_{-3.9}^{+4.7}$ &$0.32_{-0.44}^{+0.53}\pm0.03$ ($<1.5$) & $0.7$\\ \hline

\multicolumn{3}{c}{$B \to X(3872)(\to J/\psi\gamma) K$~~~~}  &  $\mathcal{B}$ ($\times 10^{-6}$)  & \\ \hline
$K^{+}$ & 18.3 & $30.0_{-7.4}^{+8.2}$& $1.78_{-0.44}^{+0.48}\pm0.12$& $4.9$ \\
$K^0$ & 14.5 & $5.7_{-2.8}^{+3.5}$ &$1.24_{-0.61}^{+0.76}\pm0.11$ ($<2.4$) & $2.4$\\  \hline

\multicolumn{3}{c}{$B \to X(3872)(\to \psi' \gamma) K$~~~~~~} & $ \mathcal{B}$ ($\times 10^{-6}$)   & \\ \hline
$K^{+}$ & 14.7 & $5.0_{-11.0}^{+11.9}$ & $0.83^{+1.98}_{-1.83}\pm 0.44$ ($<3.45$) & 0.4 \\
$K^0$ & 10.8 & $1.5_{-3.9}^{+4.8}$ & $1.12^{+3.57}_{-2.90}\pm 0.57$ ($<6.62$) & 0.3 \\ \hline \hline

\end{tabular}
\label{tab_results}
\end{center}
\end{table}

A correction for small differences in the signal detection efficiency 
calculated from signal MC and data has been applied for the lepton 
(kaon/pion) identification requirement.  Samples of $J/\psi \to \ell^+ \ell^-$ and 
$D^{*+} \to D^0(K^- \pi^+)\pi^+$  decays are used to estimate the lepton 
identification correction  and the kaon (pion) identification correction, 
respectively.  The uncertainties on these corrections are included in the
systematic error.   The errors on the PDF shapes are obtained by varying all 
fixed parameters by $\pm 1 \sigma$ and taking the change in the yield 
as the systematic error.
  To estimate the uncertainty arising from the fixed
fractions of $B\to\psi' K$ and $B\to \psi' K^*$ in the $B \to(\psi'\gamma)K$  background shape,
 we vary their   branching fractions  by $ {\pm}1 \sigma$.
The uncertainty due 
to the secondary branching fractions are  similarly taken into account. 
The uncertainty on the tracking efficiency and  the number of recorded 
$B$ meson pairs are estimated to be 1.0$\%$ 
per track  and $1.4\%$, respectively.   The uncertainty on the photon 
identification is estimated to be 2.0\% and 3.0\% for 
$B \to (J/\psi \gamma) K$ and $B \to (\psi' \gamma) K$, respectively.  
 There is some possible efficiency difference of the  selections
($E_{\gamma}, ~ \pi^0$-veto and $\cos\theta_{\rm hel}$) between
data and MC. This difference in  the $B \to(J/\psi\gamma)K$ study  
is estimated to be  $3.0\%$ using the $B^+\to\chi_{c1} K^+$ sample. 
Due to the non-availability of a proper model to generate $\chi_{c2}$  in the Evtgen simulation \cite{evtgen}, 
and the ambiguity in  the allowed $X(3872)$ $J^{PC}$  values
($1^{++}~ {\rm or}~ 2^{-+}$)~\cite{cdf_jpc}, we generate $\chi_{c2}$ 
and $X(3872)$ assuming them to be scalar, vector and tensor  particles.
We  find that 4.0 \% is the maximum 
 possible difference
in the efficiency  and include it in the systematic error.

In summary,  we observe  $X(3872)\to J/\psi \gamma$ in the $B$ decays 
and present the most precise measurement to date of the  product branching 
fraction $\mathcal{B}(B^+ \to X(3872) K^+) \mathcal{B}(X(3872) \to J/\psi\gamma)$ $=$ $(1.78^{+0.48}_{-0.44}\pm 0.12)\times 10^{-6}$. 
We also  report evidence  for $B \to \chi_{c2} K$,  
and the ratio of $\mathcal{B}(B^+ \to \chi_{c2} K^+) /$$\mathcal{B} (B^+ \to \chi_{c1} K^+)$ is   measured to be $(2.25_{-0.69}^{+0.73} \pm 0.17)\%$. 
We find no evidence for $X(3872) \to \psi'\gamma$  and give an U.L. on 
 its branching fraction as well as the following limit $R (\equiv \frac{\mathcal{B}(X(3872)\to\psi'\gamma)}{\mathcal{B}(X(3872)\to J/\psi\gamma)} ) < 2.1$ (at 90\% C.L.).  
 The $X(3872)$ state may not have a large  $c\bar{c}$ admixture  with  a $D^{*0}\overline{D}{}^0$ molecular component as was expected on the basis of  the BaBar result \cite{babarprl102}.

We thank the KEKB group for excellent operation of the
accelerator, the KEK cryogenics group for efficient solenoid
operations, and the KEK computer group and
the NII for valuable computing and SINET3 network support.  
We acknowledge support from MEXT, JSPS and Nagoya's TLPRC (Japan);
ARC and DIISR (Australia); NSFC (China); MSMT (Czechia);
DST (India); MEST, NRF, NSDC of KISTI, and WCU (Korea); MNiSW (Poland); 
MES and RFAAE (Russia); ARRS (Slovenia); SNSF (Switzerland); 
NSC and MOE (Taiwan); and DOE (USA).


\begin{thebibliography}{99}
\bibitem{wavg}
Our own average using the most recent measurements ($X(3872) \to J/\psi \pi \pi$ decay channel) from  Belle, BaBar, CDF and D$\O$ \cite{do1, bel_u, bar_u, cdf_u}.
\bibitem{psi(2S)}
 The  $\psi'$, which is sometimes designated as the $\psi(2S)$, is not a pure $S$-wave vector charmonium, rather it has a significant admixture of  $^3D_1$: $|\psi'> = \cos\phi|2^3 S_1> - \sin\phi|1^3 D_1>$.
\bibitem{mixchg}
Hereafter charge-conjugate and neutral modes are included throughout the paper unless stated otherwise.
\end{thebibliography}
\end{document}